\documentclass[12pt]{article}

\usepackage[margin=1in,nohead,a4paper,dvips]{geometry}
\usepackage{amsfonts}

\makeatletter
\def\eqnarray{%
   \stepcounter{equation}%
   \def\@currentlabel{\p@equation\theequation}%
   \global\@eqnswtrue
   \m@th
   \global\@eqcnt\z@
   \tabskip\@centering
   \let\\\@eqncr
   $$\everycr{}\halign to\displaywidth\bgroup
       \hskip\@centering$\displaystyle\tabskip\z@skip{##}$\@eqnsel
      &\global\@eqcnt\@ne\hfil$\displaystyle{\hbox{}##\hbox{}}$\hfil
      &\global\@eqcnt\tw@ $\displaystyle{##}$\hfil\tabskip\@centering
      &\global\@eqcnt\thr@@ \hb@xt@\z@\bgroup\hss##\egroup
         \tabskip\z@skip
      \cr
}
\makeatother

\def\da{\downarrow}
\def\e{\mathrm e}
\mathchardef\Delta="101
\def\iu{\mathrm i}
\def\la{\langle}
\mathchardef\Psi="109
\def\ra{\rangle}
\def\ua{\uparrow}

\newtheorem{con}{Conjecture}
\newtheorem{conp}{Conjecture}

\begin{document}

\title{Spin chains and combinatorics:\\
twisted boundary conditions}
\author{A.~V.~Razumov, Yu.~G.~Stroganov\\
\small \it Institute for High Energy Physics\\[-.5em]
\small \it 142280 Protvino, Moscow region, Russia}
\date{}

\maketitle
  
\begin{abstract}
The finite XXZ Heisenberg spin chain with twisted boundary conditions was
considered. For the case of even number of sites $N$, anisotropy parameter
$-1/2$ and twisting angle $2 \pi /3$ the Hamiltonian of the system
possesses an eigenvalue $-3N/2$. The explicit form of the corresponding
eigenvector was found for $N \le 12$. Conjecturing that this vector is the
ground state of the system we made and verified several conjectures related
to the norm of the ground state vector, its component with maximal
absolute value and some correlation functions, which have combinatorial
nature. In particular, the squared norm of the ground state vector is
probably coincides with the number of half-turn symmetric alternating sign
$N \times N$ matrices.
\end{abstract}

\leftskip 0em

\vskip 1em

The Hamiltonian of the periodic XXZ Heisenberg spin chain with the number
of sites $N$ and the anisotropy parameter $\Delta$ has the form
\[
H^{(N)} = -\sum_{i=1}^{N} \left[ \sigma_i^{x} \sigma_{i+1}^{x} +
\sigma_i^{y} \sigma_{i+1}^{y} + \Delta \, \sigma_i^z \sigma_{i+1}^z
\right].
\]
where we assume that
\[
\sigma_{N+1}^x = \sigma_1^x, \qquad \sigma_{N+1}^y = \sigma_1^y, \qquad
\sigma_{N+1}^z = \sigma_1^z.
\]
In terms of the operators $\sigma_i^+ = (\sigma_i^x + \iu \,
\sigma_i^y)/2$, $\sigma_i^- = (\sigma_i^x - \iu \, \sigma_i^y)/2$ and
$\sigma_i^z$ one has the following expression for the Hamiltonian
\begin{equation}
H^{(N)} = {} - 2 \sum_{i=1}^N \left[ \sigma_i^+ \sigma_{i+1}^- + \sigma_i^-
\sigma_{i+1}^+ \right] - \Delta \sum_{i=1}^N \sigma_i^z \sigma_{i+1}^z,
\label{1}
\end{equation}
and the boundary conditions take the form
\begin{equation}
\sigma_{N+1}^+ = \sigma_1^+, \qquad \sigma_{N+1}^- = \sigma_1^-, \qquad
\sigma_{N+1}^z = \sigma_1^z. \label{2}
\end{equation}

It was proved in paper \cite{S} that for $\Delta = -1/2$ and for an odd $N$
the Hamiltonian $H^{(N)}$ has the eigenvalue $-3N/2$, see also \cite{B}. In
our paper~\cite{RS} we analysed the explicit expressions for the
corresponding eigenvectors and made the conjecture, which was verified for
$N \le 17$, that the eigenvalue $-3N/2$ corresponds to the ground state of
the system. This is in agreement with the earlier numerical results
reported in paper \cite{ABB}. In the same paper~\cite{RS} we made and
verified for $N \le 17$ a lot of conjectures concerning the norm of the
ground state, the largest components of the ground state vector and some
correlation functions. These conjectures have a combinatorial character and
are related, in particular, to the number of alternating sign matrices
(see, for example,~\cite{Br, BP}).

Several variations of the XXZ Heisenberg chain with a special ground state
energy are also known~\cite{ABB,ABBBQ,FSZ,FSZ2}. Very recently Batchelor,
de Gier and Nienhuis considered these variations along with the
corresponding O$(n)$ loop model at $n=1$~\cite{BGN}. They found the
generalization for two of our conjectures for these systems and extended a
list of related combinatorial objects.

First of all the authors of paper \cite{BGN} reported about XXZ Heisenberg
chain with twisted boundary conditions. The Hamiltonian in this case has
again form (\ref{1}), but instead of (\ref{2}) one uses the boundary
conditions
\[
\sigma_{N+1}^+ = \e^{\iu \phi} \sigma_1^+, \qquad \sigma_{N+1}^- = \e^{-\iu
\phi} \sigma_1^-, \qquad \sigma_{N+1}^z = \sigma_1^z.
\]
Explicitly one can write
\begin{equation}
H^{(N)} = {} - 2 \sum_{i=1}^{N-1} [\, \sigma^+_i \sigma^-_{i+1} +
\sigma^-_i \sigma^+_{i+1} \,] - 2 \, \e^{-\iu \varphi} \sigma^+_N
\sigma^-_1 - 2 \, \e^{\iu \varphi} \sigma^-_N \sigma^+_1 - \Delta
\sum_{i=1}^N \sigma^z_i \sigma^z_{i+1}. \label{3}
\end{equation}
In paper \cite{ABB} on the base of numerical calculations it was
conjectured that for the special value of the twisting angle $\phi =
2\pi/3$ and $\Delta = -1/2$ the ground state energy is equal to $-3N/2$.
The proof of the existence of such an eigenvalue of the Hamiltonian was
given in paper \cite{FSZ2}. It was the case that was considered by the
authors of paper \cite{BGN}. They formulated two conjectures concerning the
sum of the components of the supposed ground state vector and the sum of
its squared components enjoying remarkable combinatorial character.

In the present paper we also consider the case $\phi = 2\pi/3$, $\Delta =
-1/2$ and study the eigenvectors of the Hamiltonian (\ref{3}) corresponding
to the eigenvalue $-3N/2$ with the main purpose to find conjectures on
correlation functions. We start with exposition of the explicit form of the
eigenvectors for $N=2,4,6$. We hope that this information will help someone
for further treating of the problem. As a matter of fact we found
explicitly the eigenvectors for $N=8,10$ and 12 as well. But they have 70,
252 and 924 components, respectively, so we decided against their
exposition. Let us fix the notations.

We use the basis in the state space formed by the normed common
eigenvectors of the operators $\sigma_i^z$. Here we associate $\ua$ with
the eigenvalue $+1$ and $\da$ with the eigenvalue $-1$. Thus, the basis
vectors are marked by words consisting of $N$ letters $\ua$ and $\da$. The
set of such words is denoted $W^{(N)}$ and we write for a general state
vector
\[
| \Psi \ra = \sum_{A \in W^{(N)}} \Psi_A | A \ra.
\]
Furthermore, we use such representation for the operators $\sigma_i^x$ and
$\sigma_i^y$ that their matrix elements coincide with usual Pauli matrices.
Finally, the eigenvector of the Hamiltonian (\ref{3}) with the eigenvalue
$-3N/2$ is denoted $|\Psi^{(N)}\ra$.

The eigenvector $|\Psi^{(2)}\ra$ has the form
\[
|\Psi^{(2)} \ra = - \frac{\iu}{2} \, (1 + \iu \, \sqrt{3}) \mid \da \ua \ra
+ \frac{\iu}{2} \, (1 - \iu \, \sqrt{3}) \mid \ua \da \ra.
\]
Let us discuss the used normalization condition. Consider the operator
\[
R = \sigma^x_1 \cdots \sigma^x_N
\]
which reverse the $z$-action projection of the spins and the antiunutary
operator $K$ of complex conjugation acting on a general state vector as 
\[
K |\Psi \ra = \sum_{A} \Psi_A^* | A \ra.
\]
It is clear that the Hamiltonian (\ref{3}) commutes with the
operator $R' = RK$\footnote{The Hamiltonian (\ref{3}) also commutes with
the product of the operator reversing the direction of the chain and the
operator $K$ and such symmetry can be also used for the fixation of
$|\Psi^{(N)}\ra$.}. Moreover, we have $R^{\prime 2} = 1$. We have found
that for even $N \le 12$ the eigenvalue $-3N/2$ is nondegenerate. It
follows
from these facts that multiplying $|\Psi^{(N)}\ra$ by an appropriate phase
factor one can satisfy the condition
\begin{equation}
R'|\Psi^{(N)}\ra = |\Psi^{(N)}\ra. \label{4}
\end{equation}
This condition is invariant under the multiplication of $|\Psi^{(N)}\ra$ by
a real number. To fix this freedom we divide $|\Psi^{(N)}\ra$ by the least
absolute value of its components. After this the vector $|\Psi^{(N)}\ra$ is
defined up to a sign which is fixed by the requirement that the sum of the
components of $|\Psi^{(N)}\ra$ be positive\footnote{Note that this sum is
real due to the condition (\ref{4}).}.

Under the above normalization conditions the vector $|\Psi^{(4)}\ra$ looks
as
\begin{eqnarray*}
|\Psi^{(4)} \ra = \frac{1}{2} \, (1 &-& \iu \sqrt{3}) \mid \da \da \ua \ua
\ra + \frac{1}{2} \, (3 - \iu \sqrt{3}) \mid \da
\ua \da \ua \ra +  \mid \da \ua \ua \da \ra \\
&+& \mid \ua \da \da \ua \ra + \frac{1}{2} \, (3 + \iu \sqrt{3}) \mid \ua
\da \ua \da \ra  + \frac{1}{2} \, (1 + \iu \sqrt{3}) \mid \ua \ua \da \da
\ra.
\end{eqnarray*}
For the case $N = 6$ we obtain
\begin{eqnarray*}
|\Psi^{(6)}\ra &=& {} - \iu
\mid \da \da \da \ua \ua \ua \ra +
\frac{1}{2} \, (\sqrt{3} - 5 \,\iu )
\mid \da \da \ua \da \ua \ua \ra \\
&+& (\sqrt{3} - 2 \, \iu)
\mid \da \da \ua \ua \da \ua \ra +
\frac{1}{2} \, (\sqrt{3} - \iu)
\mid \da \da \ua \ua \ua \da \ra +
(\sqrt{3} - 2 \,\iu)
\mid \da \ua \da \da \ua \ua \ra \\
&+& \frac{5}{2} \, (\sqrt{3} - \iu)
\mid \da \ua \da \ua \da \ua \ra +
\frac{1}{2} \, (3 \sqrt{3} - \iu)
\mid \da \ua \da \ua \ua \da \ra +
\frac{1}{2} \, (3 \sqrt{3} - \iu)
\mid \da \ua \ua \da \da \ua \ra \\
&+& \frac{1}{2} \, (3 \sqrt{3} + \iu)
\mid \da \ua \ua \da \ua \da \ra +
\frac{1}{2} \, (\sqrt{3} + \iu)
\mid \da \ua \ua \ua \da \da \ra +
\frac{1}{2} \, (\sqrt{3} - \iu)
\mid \ua \da \da \da \ua \ua \ra \\
&+& \frac{1}{2} \, (3 \sqrt{3} - \iu)
\mid \ua \da \da \ua \da \ua \ra +
\frac{1}{2} \, (3 \sqrt{3} + \iu)
\mid \ua \da \da \ua \ua \da \ra +
\frac{1}{2} \, (3 \sqrt{3} + \iu)
\mid \ua \da \ua \da \da \ua \ra \\
&+& \frac{5}{2} \, (\sqrt{3} + \iu)
\mid \ua \da \ua \da \ua \da \ra +
(\sqrt{3} + 2 \,\iu)
\mid \ua \da \ua \ua \da \da \ra + 
\frac{1}{2} \, (\sqrt{3} + \iu)
\mid \ua \ua \da \da \da \ua \ra \\
&+& (\sqrt{3} + 2 \,\iu)
\mid \ua \ua \da \da \ua \da \ra +
\frac{1}{2} \, (\sqrt{3} + 5 \, \iu)
\mid \ua \ua \da \ua \da \da \ra + 
\iu 
\mid \ua \ua \ua \da \da \da \ra.
\end{eqnarray*}

Due to the complex character of the components of the eigenvector
$|\Psi^{(N)}\ra$, our data is less than for the periodic case~\cite{RS}.
Nevertheless, we are able to formulate and verify analogues of conjectures
given in paper \cite{RS}. The first conjecture associates the eigenvalue
$-3N/2$ with the ground state.
\begin{con}
The ground state of Hamiltonian {\rm (\ref{3})} for an even $N$, $\phi =
2\pi/3$ and $\Delta = -1/2$ has the energy $E = -3N/2$ and the total spin
$z$-axes projection $S_z = 0$.
\end{con}

Consider now the maximal and the minimal absolute value of the components
of $|\Psi^{(N)}\ra$. Note that the minimal absolute value is taken, in
particular, by the ferromagnetic component $\Psi^{(N)}_{\da \ldots \da \ua
\ldots \ua}$ and the maximal absolute value is taken by the
antiferromagnetic component $\Psi^{(N)}_{\da \ua \ldots \da \ua}$. For the
absolute value of the ratio $\Psi^{(2n)}_{\da \ua \ldots \da
\ua}/\Psi^{(2n)}_{\da \ldots \da \ua \ldots \ua}$ we have the sequence
\begin{equation}
1,\quad \sqrt{3},\quad 5,\quad 14\sqrt{3},\quad 198, \quad 1573\sqrt{3},
\quad \ldots \label{5}
\end{equation}
Remind that for the periodic boundary conditions and an odd number 
of sites we have obtained~\cite{RS} that the similar ratio
$\Psi^{(2n+1)}_{\da \da \ua \ldots \da \ua}/\Psi^{(2n+1)}_{\da \da \ldots
\da \ua \ldots \ua}$ coincides with the number $A_n$ of the alternating
sign $n \times n$ matrices. The numbers $A_n$ are given by the formula 
\[
A_n = \prod_{i=0}^{n-1} \frac{(3i+1)!}{(n+i)!}.
\]
and the sequence $A_n$ goes as
\begin{equation}
1, \quad 2, \quad 7, \quad 42, \quad 429, \quad 7436, \quad \ldots
\label{6}
\end{equation}
Dividing sequence (\ref{5}) by sequence (\ref{6}), we come to
\begin{equation}
1, \quad \frac{\sqrt{3}}{2}, \quad \frac{5}{7}, \quad \frac{\sqrt{3}}{3},
\quad \frac{6}{13}, \quad \frac{11\sqrt{3}}{4\cdot 13}, \quad \ldots
\label{7}
\end{equation}
Let us now we pay the special attention to the largest prime divisors in
the denominators, 7 for $n = 3$ and 13 for $n = 5$. Their locations prompt
to consider the auxiliary sequence
\begin{equation}
1, \quad 4, \quad 4 \cdot 7, \quad 4 \cdot 7 \cdot 10, \quad 
4 \cdot 7 \cdot 10 \cdot 13, \quad 4 \cdot 7 \cdot 10 \cdot 13 \cdot 16,
\quad \ldots \label{8}
\end{equation}
and to multiply sequence (\ref{7}) by sequence (\ref{8}). This leads to
\begin{equation}
1, \quad 2 \sqrt{3}, \quad 2^2 \cdot 5,\quad \frac{2^3 \cdot 5 \cdot 7
\sqrt{3}}{3}, \quad 2^4 \cdot 3 \cdot 5 \cdot 7, \quad 2^5 \cdot 5 \cdot
7\cdot 11 \sqrt{3}, \quad \ldots \label{9}
\end{equation}
The largest prime divisors in the numerators hint at another auxiliary
sequence
\begin{equation}
1, \quad 3, \quad 3 \cdot 5, \quad 3 \cdot 5 \cdot 7, \quad 
3 \cdot 5 \cdot 7 \cdot 9, \quad 3 \cdot 5 \cdot 7 \cdot 9 \cdot 11, \quad
\ldots \label{10}
\end{equation}
Dividing sequence (\ref{9}) by sequence (\ref{10}), we arrive at
\[
1, \quad \frac{2}{\sqrt{3}}, \quad \left(\frac{2}{\sqrt{3}}\right)^2, \quad
\left(\frac{2}{\sqrt{3}}\right)^3, \quad \left(\frac{2}{\sqrt{3}}\right)^4,
\quad  \left(\frac{2}{\sqrt{3}}\right)^5, \quad \ldots
\]
As the result we obtain
\begin{con} \label{c}
The absolute value of the ratio of a component of $|\Psi^{(2n)}\ra$ with
the largest absolute value to the component with the smallest absolute
value is given by the formula
\[
\left| \Psi^{(2n)}_{\da \ua \ldots \da \ua}/\Psi^{(2n)}_{\da \ldots \da \ua
\ldots \ua} \right| = \left(\frac{2}{\sqrt{3}}\right)^{n-1} \frac{1\cdot 3
\cdot 5 \cdots (2n-3)(2n-1)}{1 \cdot 4 \cdot 7 \cdots (3n-5)(3n-2)} \, A_n.
\]
\end{con}

Consider now the squared norm of the vector $|\Psi^{(N)}\ra$:
\[
\mathcal N_{N}^2 = \la \Psi^{(N)} | \Psi^{(N)} \ra = \sum_{A \in
W^{(N)}} |\Psi^{(N)}_A|^2.
\]
Fixing the normalization in the way discussed above we see that the
sequence formed by the numbers $\mathcal N_{2n}^2$ goes as
\begin{equation}
2, \quad 10, \quad 140, \quad 5544, \quad 622908, \quad 198846076, \quad
\dots \label{11}
\end{equation}
For the case of periodic boundary conditions the corresponding sequence of
squared norms $\mathcal N_{2n+1}^2$ is
\begin{equation}
1, \quad 3, \quad 25, \quad 588, \quad 39204, \quad 7422987, \quad
\ldots \label{13}
\end{equation}
and we conjectured in paper~\cite{RS} that
\[
\mathcal N_{2n+1}^2 = \left(\frac{3}{4}\right)^n \left[\frac{ 2 \cdot 5
\cdots (3n-1)} {1 \cdot 3 \cdots (2n-1)}\right]^2  A_n^2.
\]
Dividing sequence (\ref{11}) by the sequence formed by the numbers
$A_n^2$, we obtain
\[
2, \quad \frac{5}{2}, \quad \frac{2^2\cdot 5}{7}, \quad \frac{2\cdot
11}{7}, \quad \frac{2^2\cdot 11}{13}, \quad \frac{11\cdot 17}{2^2\cdot 13},
\quad \ldots
\]
Using the same method as for conjecture \ref{c}, we come to the following
analogue of conjecture 3 formulated in paper \cite{RS}.
\begin{con} 
The squared norm of the vector $|\Psi^{(2n)}\ra$ is given by
the formula
\[
\mathcal N_{2n}^2 = \frac{2 \cdot 5 \cdots (3n-1)}{1 \cdot 4 \cdots (3n-2)}
\, A_n^2.
\]
\end{con}

Let us remark that joining the sequences (\ref{11}) and (\ref{13}) we come
to the sequence
\[
1, \quad 2, \quad 3, \quad 10, \quad 25, \quad 140, \quad 588, \quad 5544,
\quad 39204, \quad 622908, \quad 7422987, \quad \ldots
\]
This is the sequence formed by the number of half-turn symmetric
alternating sign $N \times N$ matrices $A^{HT}_N$ (see \cite{R, K} and
references therein). Therefore we can combine our conjecture 3 and
conjecture 3 of paper \cite{RS} into the statement that $\mathcal N_N^2 =
A^{HT}_N$.

Batchelor, de Gier and Nienhuis discovered another analogue of
the conjecture 3  of paper \cite{RS}.
\setcounter{conp}{\thecon}
\addtocounter{conp}{-1}
\begin{conp}[Batchelor, de Gier and Nienhuis \cite{BGN}]
The sum of the squared components of the vector $|\Psi^{(2n)}\ra$ is given
by
\[
\sum_{A \in W^{(2n)}} [ \Psi^{(2n)}_A ]^2 = A_n^2.
\]
\end{conp}
The same authors gave the analogue of conjecture 4 from paper \cite{RS}.
\begin{con}[Batchelor, de Gier and Nienhuis \cite{BGN}]
The sum of the components of the vector $|\Psi^{(2n)}\ra$ is given
by
\[
\sum_{A \in W^{(2n)}} \Psi^{(2n)}_A = 3^{n/2} A_n.
\]
\end{con}

Note that Hamiltonian (\ref{3}) has a modified shift invariance.
Indeed, it commutes with the operator
\[
T' = T \exp \left[ \frac{\iu \phi}{2} \left( \sigma^z_N -
\frac{1}{N}\sum_{i=1}^N \sigma_i^z \right) \right],
\]
where $T$ is the operator of the right shift by a site. Our choice of the
operator of modified shift invariance is dictated by the requirement
$T^{\prime N} = 1$. Note that the operator $T'$ commutes with the operator
$R'$. It follows from these facts and from requirement (\ref{4}) that in
the case under consideration either $T'|\Psi^{(N)}\ra = |\Psi^{(N)}\ra$ or
$T'|\Psi^{(N)}\ra = - |\Psi^{(N)}\ra$. It appears that for even $N \le 12$
one has
\begin{equation}
T'|\Psi^{(N)}\ra = |\Psi^{(N)}\ra. \label{12}
\end{equation}
It is natural to suppose that it is so for any even $N$. It can be deduced
from (\ref{12}) that for any $k = 0, 1, \ldots$
\[
\sum_{A \in W^{(N)}} [\Psi^{(N)}_{A}]^{3(2k+1)} = 0.
\]

Let us proceed to the consideration of correlation functions. For any
operator $O$ we denote
\[
\la O \ra_N = \frac{\la \Psi^{(N)}| O |\Psi^{(N)} \ra}{\la \Psi^{(N)} |
\Psi^{(N)} \ra}.
\]
It is evident that $\la \sigma^z_{i_1} \ldots \sigma^z_{i_k} \ra_{2n}$ does
not depend on the order of the operators $\sigma^z_i$. From (\ref{4}) we
obtain
\[
\la \sigma^z_{i_1} \ldots \sigma^z_{i_{2k+1}} \ra_{2n} = 0
\]
for any $k = 0, 1, \ldots$, and the equality (\ref{12}) implies that
\[
\la \sigma^z_{i_1} \ldots \sigma^z_{i_k} \ra_{2n} = \la \sigma^z_{i_1+1}
\ldots \sigma^z_{i_k+1} \ra_{2n}.
\]
To formulate our conjectures on correlation functions introduce the
operators 
\[
\alpha_i = (1 + \sigma^z_i)/2.
\]
The analogue of conjecture 5 of paper \cite{RS} is

\begin{con}
\[
\la \alpha_i \, \alpha_{i+2} \ra_{2n} = \frac{71 n^4 - 19 n^2 + 20}{16(4
n^2 - 1)^2}.
\]
\end{con}

\noindent Unlike the periodic case, this conjecture is a simple consequence
of

\begin{con}
There is a simple formula for the correlation functions called the
Probabilities of Formation of Ferromagnetic String {\rm \cite{KIB}}:
\[
\frac{\la \alpha_1 \, \alpha_2 \ldots \alpha_{k-1} \ra_{2n}}{\la \alpha_1
\, \alpha_2 \ldots \alpha_k \ra_{2n}} = \frac{(2k - 2)! \, (2k - 1)! \, (2n
+ k - 1)! \, (n - k)!}{(k - 1)! \, (3k - 2)! \, (2n - k)! \, (n + k - 1)!}.
\]
\end{con}

Note that the conjectured formulas for correlation functions have the same
thermodynamic limit both for the twisted and  periodic boundary conditions
\cite{RS}.

{\it Acknowledgments} The work was supported in part by the Russian
Foundation for Basic Research.

\end{document}